\documentstyle[11pt,mrs2001,epsfig]{article}
\newcommand{\apg}{\:^{>}_{\sim}\:}
\newcommand{\apl}{\:^{<}_{\sim}\:}
\newcommand{\etal}{et al.}
\begin{document}
\title{The $z\sim 1.2$ Galaxy Luminosity Function from The LCIR Survey}

\author{H.-W. Chen$^1$, P. J. McCarthy$^1$, R. O. Marzke$^2$, 
R. G. Carlberg$^3$, A. E. Firth$^4$, S. E. Persson$^1$,
R. G. McMahon$^4$, O. Lahav$^4$, P. Martini$^1$, R. S. Ellis$^5$, 
R. G. Abraham$^3$, A. Oemler$^1$, R. S. Somerville$^4$}
\affil{$^1$Carnegie Observatories, 813 Santa Barbara St, Pasadena, CA 91101, 
U.S.A.}
\affil{$^2$Department of Astronomy and Physics, San Francisco State University,
San Francisco, CA 94132-4163, U.S.A.}
\affil{$^3$Department of Astronomy, University of Toronto, Toronto ON, M5S~3H8
Canada}
\affil{$^4$Institute of Astronomy, Cambridge CB3 OHA, England, UK}
\affil{$^5$Department of Astronomy, Caltech 105-24, Pasadena, CA 91125, U.S.A.}

\begin{abstract}

  We present results from the Las Campanas Infrared Survey, designed to 
identify a statistically significant sample of $z\apg 1$ galaxies using 
photometric redshift techniques.  Here we summarize the design and strategies 
of the survey and present the first estimate of the galaxy luminosity function 
at $z\apg 1$ based on $H$-band selected galaxies identified in our survey. 
Results of number count studies and luminosity function measurements indicate 
that most early-type galaxies were already in place by $z\sim 1.2$ with a 
modest space density evolution and a mild luminosity evolution over that
expected from passive evolution.
\end{abstract}

\section{Introduction}

  Evolved high-redshift galaxies hold the key to addressing two fundamental
issues in observational cosmology: First, when did the bulk of star formation 
and mass assembly occur?  Second, how does the space density of galaxies evolve
with time?  Because different galaxy formation scenarios have distinct 
predictions for the space densities and masses of galaxies at redshifts beyond 
one, comparisons of statistical properties of evolved high-redshift galaxies 
and local ellipticals may provide a direct means of discriminating between 
competing galaxy formation scenarios (\cite{kc98}).  Previous studies of 
evolved galaxies have yielded inconsistent measurements of their space density
(see \cite{cmm01} for a list of references) due to a strong field-to-field 
variation in the surface density (\cite{d00,mcmc01a,mcmc01b,fsml01}) and 
various selection biases (\cite{ty97}).  Evolved galaxies may be characterized
by their intrinsically red colors due to the lack of ongoing star formation 
that provides most of the UV light in typical galaxy spectral energy 
distributions, but confusions may arise because of the presence of dusty 
star-forming galaxies that exhibit similar colors.

  We have been conducting the Las Campanas Infrared (LCIR) Survey in the past 
two years.  It is a deep near-infrared and optical imaging survey of distant 
galaxies over one square degree of sky at high galactic latitudes (\cite{mmp99,
mcmc01a,mcmc01b,cmm01}).  This program utilizes one of the largest 
near-infrared cameras available (CIRSI;\cite{b98}) that produces an image of 
$13' \times 13'$ contiguous field of view in a sequence of four pointings.  The
survey is designed to identify a large number of red galaxies at $z>1$ using 
photometric redshift techniques.  The primary objectives of the program are (1)
to examine the nature of the red galaxy population and identify evolved 
galaxies at $z>1$, (2) to study the space density and luminosity evolution of 
evolved galaxies at $z\apl 2$, and (3) to measure the spatial clustering of 
evolved high-redshift galaxies (\cite{mcmc01a,mcmc01b,fsml01}), thereby 
inferring merging rates of these galaxies for constraining theoretical models.
Here we summarize the survey design and present initial results from comparing 
optical and near-infrared colors, number density and luminosity evolution 
between galaxies at different redshifts.

\section{Current Status of the Survey}


  We have completed an $H$-band imaging survey in six distinct high galactic
latitude fields, covering 1.1 square degrees of sky, to a mean $5\,\sigma$ 
detection limit in a four arcsecond diameter aperture of $H\sim 20.8$.  We have
also obtained complementary optical images of these fields to consistent depths
required to identify red galaxies with $R-K_s \apg 5$ or $I-H \apg 3$ 
(\cite{mmp99,cmm01}).  Details of the $H$-band survey status and sample 
photometric catalogs are presented in \cite{cmm01}.  The current survey depths
allow us to identify evolved galaxies that are more than one magnitude fainter
than an $L*$ elliptical galaxy at $z\sim 1$.  


  We have initiated a $K_s$-band imaging survey of the six fields using a 
re-imaging camera developed by Persson \etal\ (\cite{p01}) that provides an
effective cold stop, extending the operational range of CIRSI to the $K_s$ 
band.  The primary goal of the $K_s$-band survey is to extend the redshift 
range probed by the $H$-band survey, in order for us to identify sub-$L_*$ 
galaxies at $z\apg 1.5$.  We have observed $\approx$ 320 square arcminutes of 
sky to a mean $5\,\sigma$ detection limit in a four arcsecond diameter aperture
of $K_s\sim 20$.  More than 3,300 galaxies are identified in the $K_s$-band 
survey so far, of which $\apg 550$ have $R - K_s \apg 5$.

\section{Photometric Redshift Analysis and Optical and Near-infrared Colors}

\begin{figure}
\plottwo{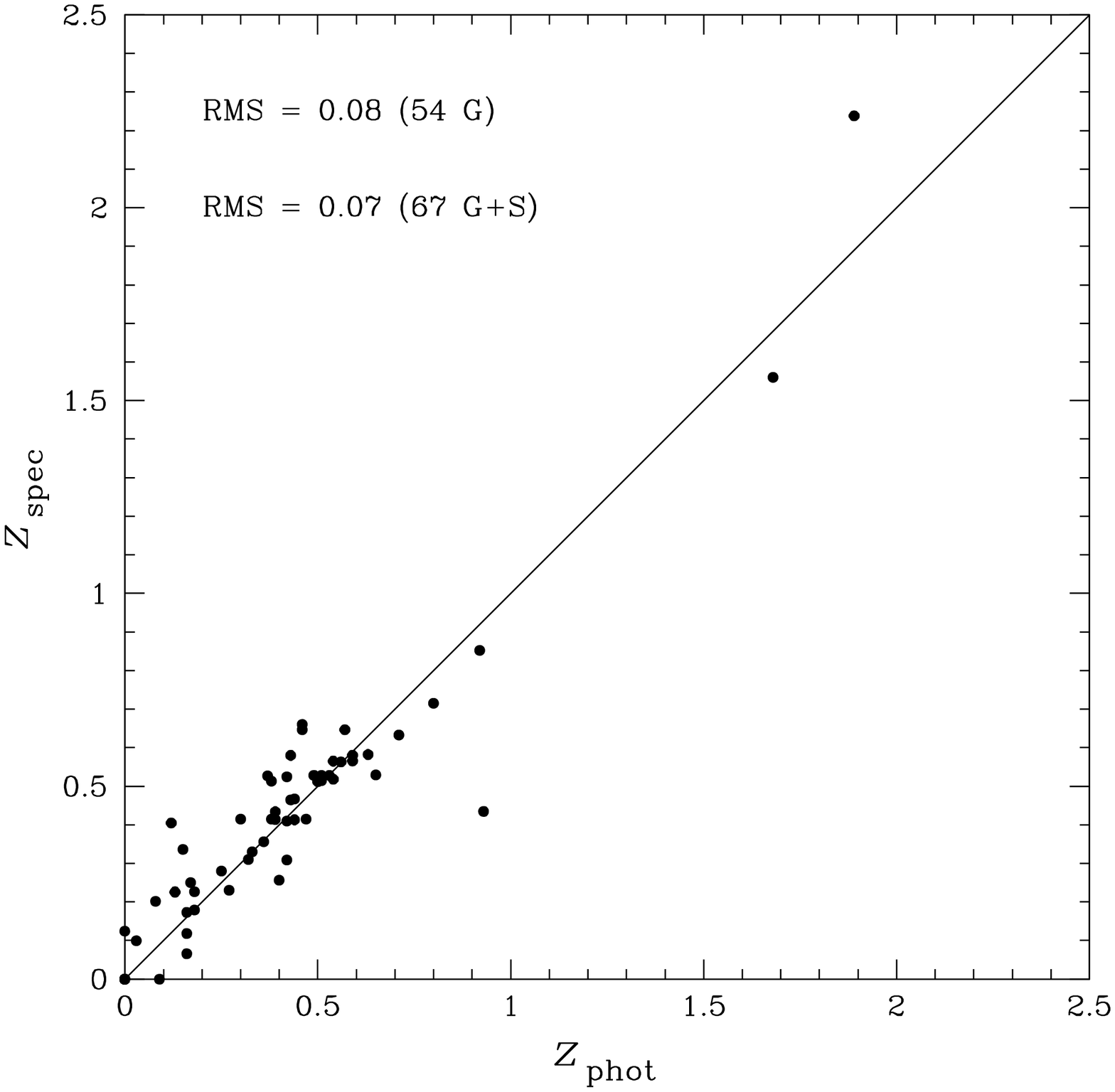}{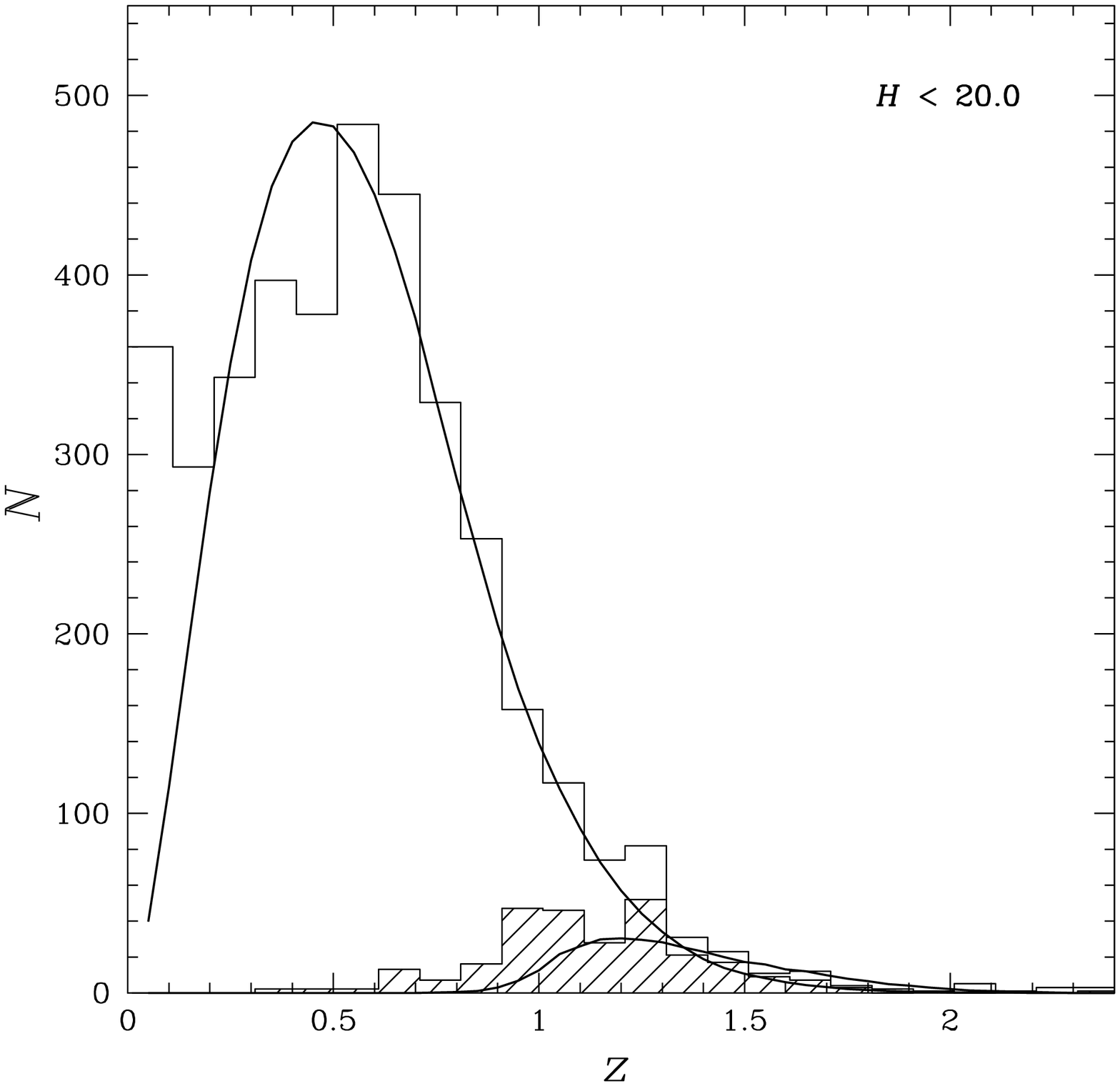}
\caption{(a) Comparison of photometric redshifts $z_{\rm phot}$ and 
spectroscopic redshifts $z_{\rm spec}$ for 13 stars and 54 galaxies observed 
spectroscopically in the HDFS.  (b) Redshift histogram of the $H$-band 
detected galaxies in the HDFS region.  The shaded part indicates the redshift 
distribution of the $I-H\apg 3$ galaxies.  Solid curves indicate the predicted
redshift distribution based on the model that best fits the differential 
surface density measurements.}
\end{figure}

  Over the past several years, it has been demonstrated that distant galaxies 
may be accurately and reliably identified using photometric redshift techniques
that incorporate optical and near-infrared broad-band photometry (\cite{cs97,
sly97,fly99}).  For galaxies lacking strong emission or absorption features, we
are still able to estimate redshifts based on the presence/absence of spectral 
discontinuities in broad-band optical and near-infrared colors.  To identify 
distant red galaxies, we have adopted the photometric redshift technique 
developed originally by Lanzetta and collaborators (\cite{lyf96,fly99}), and 
modified the program to account for the large number of stars that appear in 
wide-field surveys by including a suite of stellar templates---from early-type
OB stars to late-type L and T dwarfs.  

  Comparison of photometric redshifts $z_{\rm phot}$ and spectroscopic 
redshifts $z_{\rm spec}$ at redshifts $0 \leq z < 2.5$ for 67 objects with
spectroscopic redshifts in the Hubble Deep Field South (HDFS) shows that the 
photometric redshifts are both accurate and reliable, with an RMS dispersion 
between the photometric and spectroscopic redshifts of $\Delta z/(1+z) \approx 
0.1$ (Figure 1a).  The accuracy and reliability of photometric redshift 
measurements allow us to study the galaxy two-point correlation function and 
the galaxy luminosity function as a function of redshift at $z \apl 2.5$ using 
the wide-field survey data.  The lower precision of photometric redshift 
measurements may be compensated by the large sample size.  We show in Figure 1b
the redshift histogram of the $H$-band detected galaxies and of those with $I-H
\apg 3$ identified in the HDFS region.  The curves indicate the predicted 
redshift distribution of the total and red populations based on the model that
best fits the differential surface density measurements as described below in 
\S\ 4.1.  The $I-H\apg 3$ color criterion is determined based on predictions of
various evolution scenarios to select evolved galaxies at $z\apg 1$.  It is 
analogous to a color selection of $R-K_s \apg 5$ (\cite{cmm01}).  Figure 1b 
shows that more than 6\% of the $H$-band detected galaxies have $I-H \apg 3$ 
and $z\apg 1$.

  We present the $V-I$ vs. $I-H$ color--color distribution of the $H$-band 
detected objects in Figure 2a and the $V-I$ vs. $J-K_s$ color--color 
distribution of the $K_s$-band detected objects in Figure 2b, along with 
predictions based on different star formation histories.  Different symbols 
represent objects at different redshifts according to the results of our 
photometric redshift analysis.  The general agreement between distributions of 
various model predictions and measurements of galaxies in our catalog, and the 
clear separation of the stellar sequence (which is most obvious in the $V-I$ 
vs. $J-K_s$ plot) suggest that most stars have been accurately identified in 
our survey fields.  The large scatter in the $V-I$ colors (corresponding to the
rest-frame UV colors at $z\apg 1$) spanned by the galaxies of red $I-H$ or 
$J-K_s$ colors reveals a wide range of star formation histories for which our 
simple models provide only an idealized description.  In particular, the large 
scatter in the $V-I$ colors for galaxies of $I-H\apg 3$ or $J-K_s\apg 1.9$ 
(indicative of redshifts $z\apg 1$) suggest a significant amount of ongoing 
star formation in these galaxies.  In addition, the small fraction of galaxies 
with $I-H\apg 3$ and $V-I\apg 3$ (or $J-K\apg 2$ and $V-I\apg 3$) suggests that
there are few objects that may be adequately characterized as pure passively
evolving systems formed at high redshifts.

\begin{figure}
\plottwo{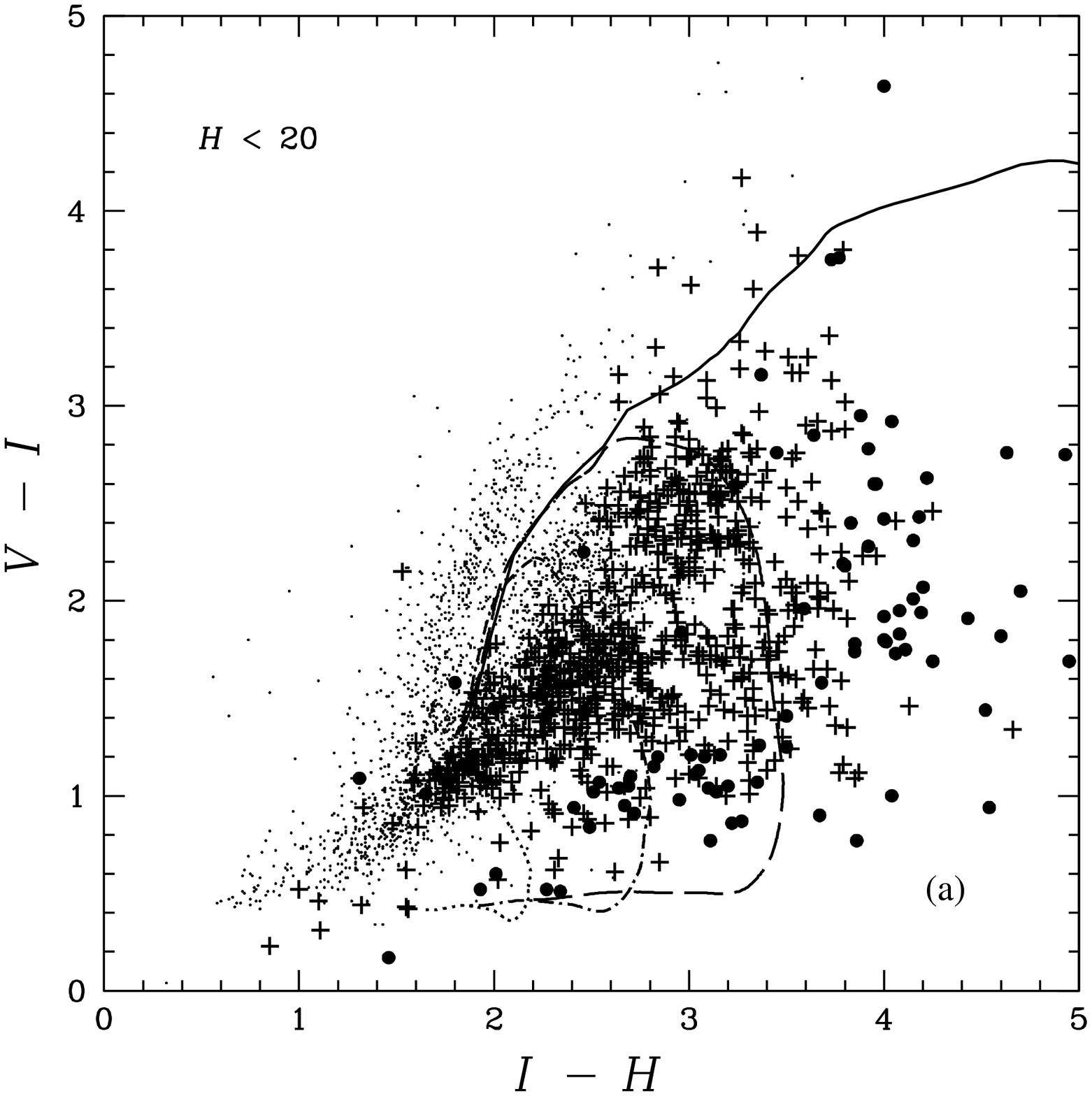}{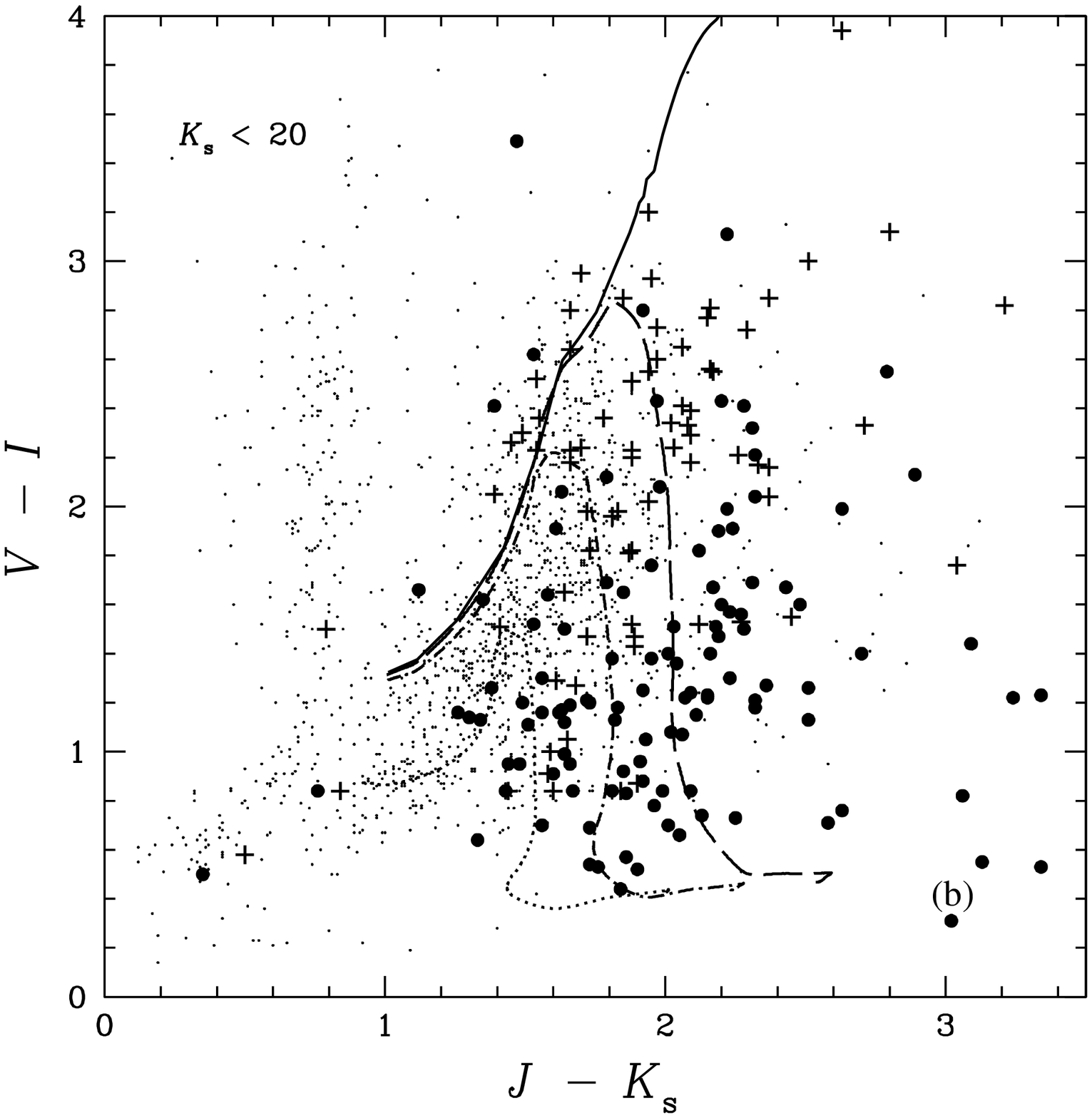}
\caption{The $V-I$ vs. $I-H$ colors (a) and $V-I$ vs. $J-K_s$ (b) colors for 
objects identified in one of the survey fields.  Different symbols represent 
objects at different redshifts according to the results of our photometric 
redshift analysis: $z\apl 0.75$ (dots), $0.75\apl z \apl 1.0$ (circles), $1.0
\apl z\apl 1.5$ (crosses).  The model curves show the loci of evolving 
population models with various star formation laws: a single burst with $z_f = 
30$ (solid curve), exponentially declining with $\tau = 1 \& 2$ Gyr (dashed and
dash-dotted curves), and continuous (dotted curve).}
\end{figure}

\section{Statistical Properties of Red Galaxies}

\subsection{Galaxy Number-Magnitude Relation}

  Figure 3a shows the number-magnitude relation of the $H$-band detected 
galaxies (open squares) and the $H$-band detected red galaxies (closed points)
along with model predictions (curves) and previous measurements (open points).
The number-magnitude relation of the $H$-band detected galaxies may be 
described using a local $K$-band luminosity function (\cite{g97}) with pure 
luminosity evolution expected from a 1 Gyr exponentially declining star 
formation rate model.  To model the number-magnitude relation of the red 
population, we scale the luminosity function according to the population ratio 
of elliptical galaxies observed in the local universe and identify red 
galaxies as those brighter than $M_*-1$ with a faint-end slope $\alpha=1$, 
$M_*({\rm red})=M_*-0.2$, and a space density $\phi_*({\rm red})=0.15\phi_*$. 
We also consider a number density evolution characterized by $(1+z)^{-p}$ with
$p = 0$, 0.5, and 1.0 (solid curves from top to bottom).  Our measurements are 
consistent with previous surface density measurements both for the total and 
the red populations.  Figure 3a shows that a luminosity evolution model with a
mild space density evolution may explain the faint number counts.

  The much steeper bright-end slope of the number-magnitude relation of the red
galaxies indicates that most of the foreground ($z\apl 1$) galaxies have been 
effectively excluded from the red population and that the slope reflects the 
shape of the bright-end luminosity function of the red population.  In 
addition, we find that red galaxies with $I-H \apg 3$ constitute $\approx$ 20\%
of the $H$-band detected galaxies at $H\apl 21$, but only $\approx$ 2\% at $H
\apl 19$.  A strong field-to-field variation in the surface density of the red 
population is directly observed in our survey, which is expected according to
the results of angular clustering analysis of these red galaxies (\cite{d00,
mcmc01b,fsml01}).  We estimate a mean surface density of $\approx 1$ 
arcmin$^{-2}$ with an rms dispersion of $\approx 0.4$ arcmin$^{-2}$ for red 
galaxies brighter than $H=20.5$.  Finally, the consistent slopes between the 
red subsamples suggest similar underlying luminosity functions.

\begin{figure}
\plottwo{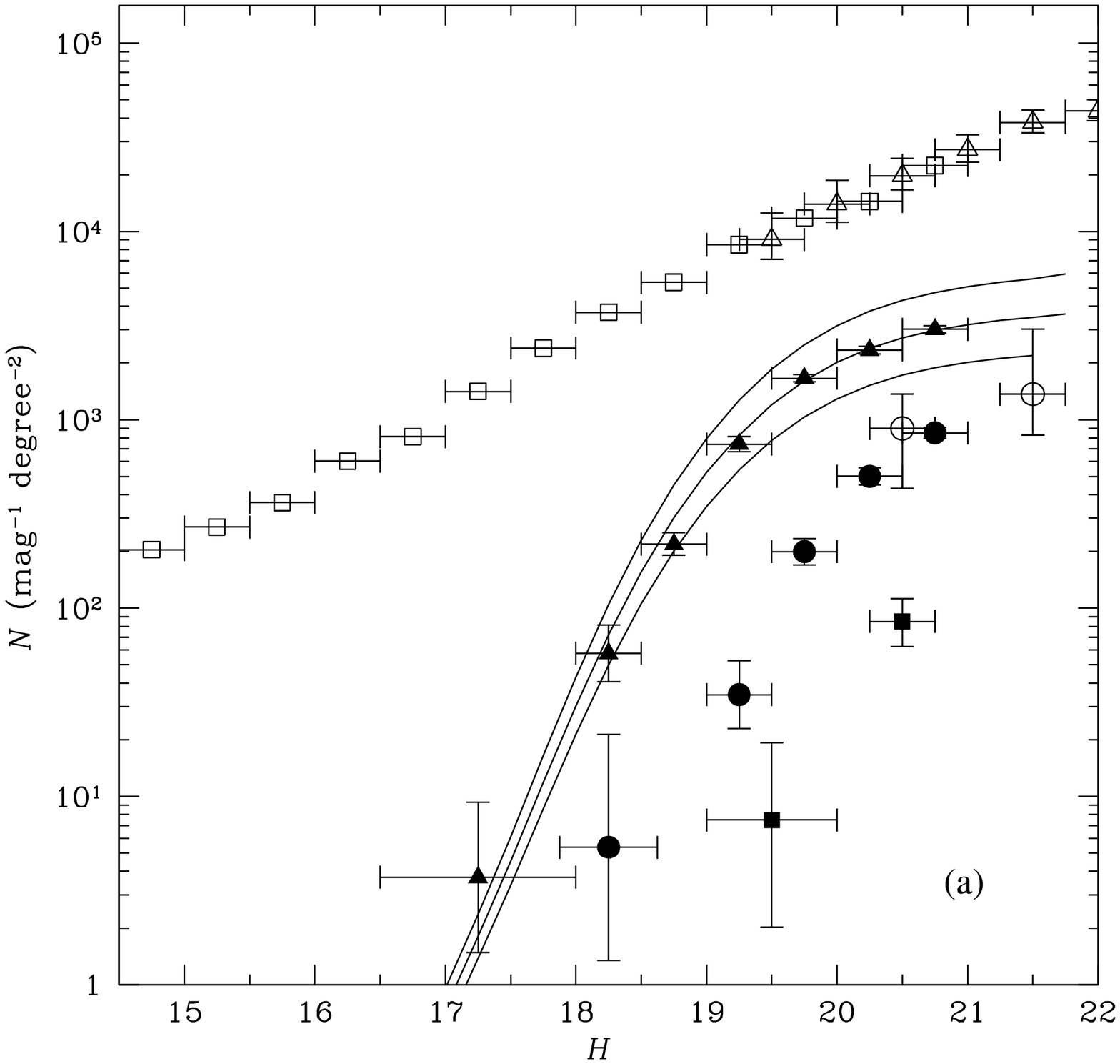}{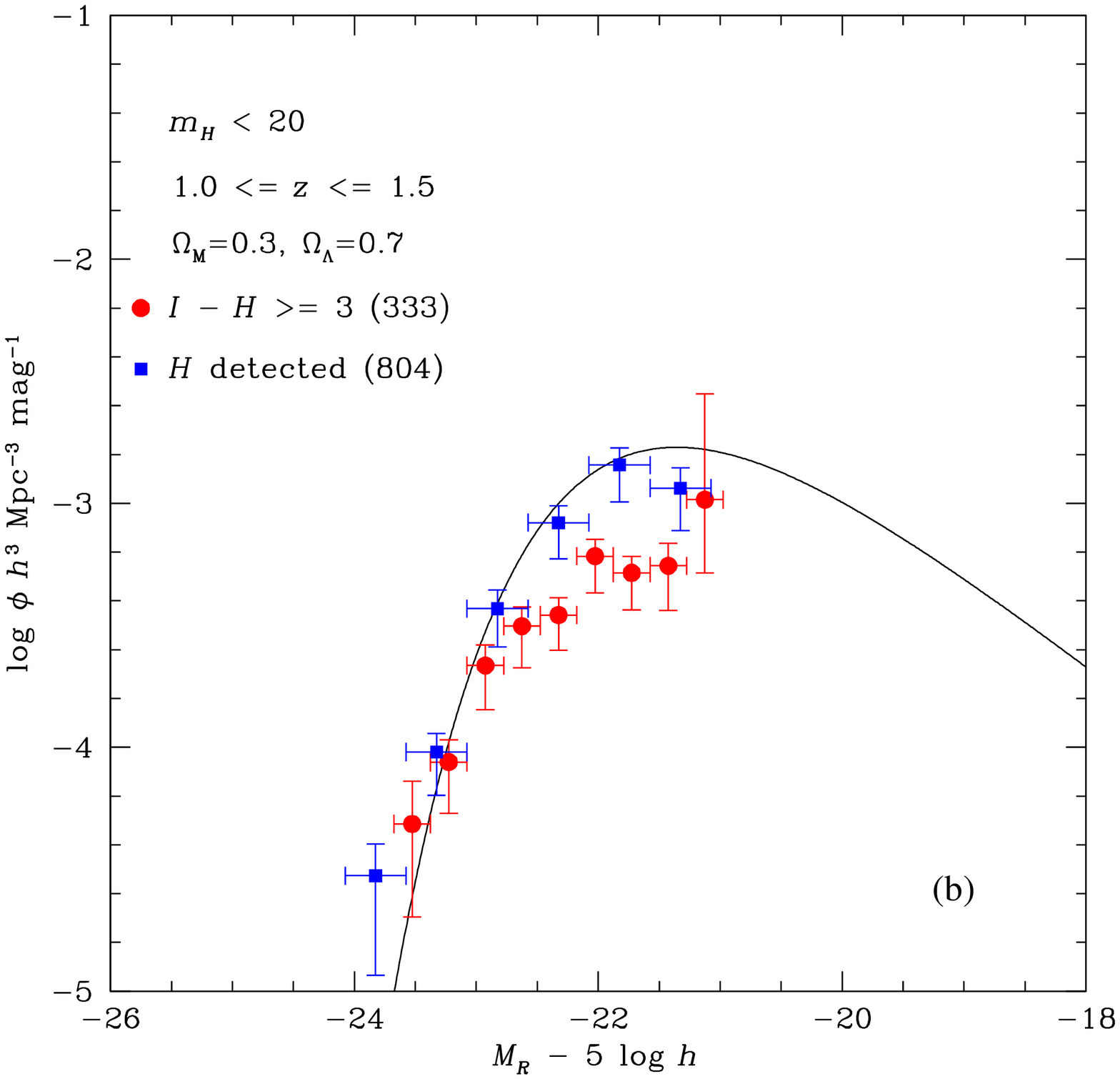}
\caption{(a) The total (open squares) and red (closed points) galaxy 
differential surface density as a function of $H$-band magnitude in four of the
survey fields.  The open triangles are NICMOS measurements presented in 
\cite{y98}.  The closed triangles, circles, and squares are the surface density
measurements for $I-H\apg 3$, 4 and 5, respectively.  The error bars show 
the 95\% confidence interval assuming Poisson statistics.  The open circles at
$H = 20.5$ and 21.5 are from the $R - H \apg 5$ NICMOS sample in \cite{y00}.  
The curves show model predictions derived from a simple evolving population 
model as described in the text.  (b) Rest-frame $R$-band luminosity function 
measurements of the $H$-band selected galaxies identified in two of the survey 
fields at redshifts between $z=1$ and 1.5.  The solid curve is a scaled CNOC2 
luminosity function of color selected early-type galaxies at $z\sim 0.3$.  
There are 804 $H$-band detected galaxies with $m_H \apl 20$ and at $1.0\apl z
\apl 1.5$, 333 of which have $I-H\apg 3$.}
\end{figure}

\subsection{Galaxy Luminosity Function}

Various deep redshift surveys have yielded consistent measurements of the
luminosity function for galaxies at $z\apl 0.75$ (see e.g.\ \cite{m98,l99}), 
but it becomes exceedingly difficult beyond this redshift range both because 
galaxies become even fainter and because bright sky lines make accurate 
spectroscopic redshift identifications challenging.  Despite a lower precision,
the large number of distant galaxies expected from a wide-field multi-color 
survey allow us to obtain statistically significant estimates of various galaxy
properties using photometric redshift measurements.  Here we present the first
measurements of the galaxy luminosity function at $1.0\apl z\apl 1.5$ based on
photometric redshifts. 

  Figure 3b shows the rest-frame $R$-band luminosity functions of the total 
$H$-band selected sample and galaxies with $I-H\apg 3$ in two of our survey 
fields.  At $z\sim 1.2$, the observed-frame $H$-band roughly corresponds to the
rest-frame $R$-band, therefore the calculation does not rely heavily on the 
adopted templates to determine the $k$-correction.  We calculate the luminosity
function in each magnitude bin using the $1/V_{\rm max}$ method, where $V_{\rm
max}$ is estimated as the comoving volume accessible by a galaxy in the sample
based on its absolute magnitude.  Errors are calculated using a bootstrap 
method, including sampling errors and measurement uncertainties in photometric
redshift, galaxy photometry, and colors.  Comparison of the luminosity 
functions of the total and red populations shows that red galaxies dominate the
bright galaxy population at $z\apg 1$.  Specifically, the space densities of 
the two populations are statistically identical at the bright end.  The space 
densities of red galaxy and galaxies with bluer $I-H$ colors become comparable
at around $M_*$.  Finally, our $H$-band survey is insensitive to sub-$L*$ 
galaxies, and so we cannot constrain the faint-end slope of the luminosity 
function.

  To assess luminosity function evolution, we compare our measurements with
those presented in \cite{l99}.  The solid curve in Figure 3b shows a scaled 
CNOC2 luminosity function of color selected early-type galaxies at $z\sim 0.3$.
First, we estimate a luminosity evolution using the same 1-Gyr exponentially 
declining star formation model and find that $M_{R_*} (z\sim 1.2) = M_{R_*}(z
\sim 0.3) - 0.8$.  Next, we find that a scaling factor in the space density 
$\phi_*(z=1.2) = 0.55\,\phi_*(z=0.3)$ is necessary to match the curve with our
measurements, consistent with the results of the number-magnitude analysis.  
However, the degree to which the CNOC2 and LCIRs samples probe the same galaxy
population remains uncertain.  In summary, the results of number count studies 
and luminosity function measurements indicate that most early-type galaxies 
were already in place by $z\sim 1.2$ with a modest space density evolution and
a mild luminosity evolution over that expected from passive evolution.

\acknowledgements{This research was supported by the National Science 
Foundation under grant AST-9900806.  H.-W. Chen acknowledges partial support
from an International Travel Grant provided by the American Astronomy Society.
The CIRSI camera was made possible by the generous support of the Raymond and 
Beverly Sackler Foundation.}

\vfill
\end{document}